\documentclass[twocolumn,aps,prb,amssymb,amstext,showpacs]{revtex4}

\usepackage{epsfig}
\usepackage{amsfonts,amssymb,amsmath}

\newcommand{\half}{\mbox{$\frac{1}{2}$}}
\newcommand{\beq}{\begin{equation}}
\newcommand{\eeq}{\end{equation}}

\newcommand{\R}{\mathbb{R}}

\newcommand{\e}{\epsilon}

\newcommand{\m}{{\bf m}}
\newcommand{\field}{{\bf H}}
\newcommand{\xhat}{\hat{{\bf x}}}
\newcommand{\yhat}{\hat{{\bf y}}}
\newcommand{\ud}{\mathrm{d}}

\newcommand{\Dwid}{d_0}

\newcommand{\sech}{\operatorname{sech}}

\newcommand\braket[2]{\left\langle #1 \right.\left|#2\right\rangle}
\newcommand\me[3]{\left\langle #1 \right | #2 \left|#3\right\rangle}

\begin{document}

\title{Stability of precessing domain walls in ferromagnetic
  nanowires}

\author{Yan Gou$^1$, Arseni Goussev$^{1,2}$, JM Robbins$^1$, Valeriy Slastikov$^1$}

\affiliation{$^1$School of Mathematics, University of Bristol,
  University Walk, Bristol BS8 1TW, United Kingdom\\ $^2$Max Planck
  Institute for the Physics of Complex Systems, N{\"o}thnitzer
  Stra{\ss}e 38, D-01187 Dresden, Germany}

\date{\today}

\begin{abstract}
  We show that recently reported precessing solution of
  Landau-Lifshitz-Gilbert equations in ferromagnetic nanowires is
  stable under small perturbations of initial data, applied field and
  anisotropy constant.  Linear stability is established analytically, while nonlinear stability is verified numerically.
\end{abstract}


\maketitle


\section{Introduction} 

The manipulation and control of magnetic domain walls (DWs) in
ferromagnetic nanowires has recently become a subject of intense
experimental and theoretical research. The rapidly growing interest in
the physics of the DW motion can be mainly explained by a promising
possibility of using DWs as the basis for next-generation memory and
logic devices
\cite{Allwood05,Cowburn07,Parkin08,Hayashi08,Thomas10}. However, in
order to realize such devices in practice it is essential to be able
to position individual DWs precisely along magnetic
nanowires. Generally, this can be achieved by either applying external
magnetic field to the nanowire, or by generating pulses of
spin-polarized electric current. The current study is concerned with
the former approach.

Even though the physics of magnetic DW motion under the influence of
external magnetic fields has been studied for more than half a century
\cite{LandoLifshitz35,Gilbert55,SchryerWalker74,Kosevich90}, current
understanding of the problem is far from complete and many new
phenomena have been discovered only recently
\cite{Hickey08,Wang09,Wang09b,Sun10,our_PRL}. In particular, a new
regime has been reported \cite{Sun10,our_PRL} in which rigid profile
DWs travel along a thin, cylindrically symmetric nanowire with their
magnetization orientation precessing around the propagation axis. In
this paper we address the stability of the propagation of such
precessing DWs with respect to perturbations of the initial
magnetization profile, some anisotropy properties of the nanowire, and
applied magnetic field.

Let $\mathbf{m}(x) = (\cos\theta(x), \sin \theta(x) \cos \phi(x), \sin \theta(x) \sin \phi(x))$ denote the 
magnetization along a one-dimensional wire.
With easy magnetization axis along $\mathbf{\hat x}$ and hard axis
along $\mathbf{\hat y}$, the micromagnetic energy is given by \cite{SS}
\begin{multline}
\label{ eq: E}
E(\mathbf{m}) = \half \int \left(A\mathbf{m'}^2 +K_1(1 - m_1^2) + K_2 m_2^2\right)\,dx\\= 
\half\int\left( A {\theta'}^2 + \sin^2\theta (A{\phi'}^2+ K_1 + K_2\cos^2\phi)\right) \,dx
\end{multline}
where $A$ is the exchange constant and $K_1$, $K_2$ the anistropy
constants.  Here and in what follows, integrals are taken between
$-\infty$ and $\infty$ (for the sake of brevity, limits of integration
will be omitted).

We consider here the case of uniaxial anisotropy,  $K_2 = 0$.  Minimizers of $E$ subject to   
the boundary conditions
\begin{equation}
\label{ eq: bcs}
\lim_{x\rightarrow \pm \infty} \mathbf{m}(x) = \pm\mathbf{\hat x}, 
\end{equation}
describe optimal profiles for a domain wall separating two magnetic domains with opposite orientation.  The optimal profiles satisfy the Euler-Lagrange equation 
\begin{equation}\label{eq: Euler-Lagrange}
\mathbf{m}\times\mathbf{H} = 0,
\end{equation}
where
\begin{equation}
\label{ eq: H}
\mathbf{H} = -\frac{\delta E}{\delta \mathbf{m}} = A\mathbf{m''} + K_1( \mathbf{m}\cdot \mathbf{\hat x}) \mathbf{\hat x}
= -e_0 \mathbf{m} + e_1 \mathbf{n} + e_2 \mathbf{p}.
\end{equation}
Here  $\mathbf{m}$,  $\mathbf{n} = \partial \mathbf{m}/\partial \theta$ and $\mathbf{p} = \mathbf{m}\times\mathbf{n}$ form an orthonormal frame, and the components of $\mathbf{H}$ in this frame are given by
\begin{align}
\label{eq: e_0 - e_2}
   e_0  &=  A{\theta'}^2 + \sin^2\theta (K_1 + A{\phi'}^2)    \nonumber \\
    e_1 &= A\theta'' - \frac12 \sin 2\theta (K_1 + A{\phi'}^2),\nonumber\\
    e_2 &= A\sin\theta \phi'' + 2A\cos\theta \theta' \phi'.
\end{align}
In terms of these components,
the energy Eq.~\eqref{ eq: E} (with $K_2 = 0$) is given by
\begin{equation}
\label{ eq: energy in terms of e_0}
E(\mathbf{m}) = \half \int e_0 \,dx,
\end{equation}
and the Euler-Lagrange equation becomes $e_1 = e_2 = 0$.

While the energy $E$ is invariant under translations along and rotations about the $x$-axis, the optimal profiles cannot be so invariant (because of the boundary conditions).  Instead, the optimal profiles form a two-parameter family  obtained  by applying translations, denoted $T(s)$, and rotations, denoted $R(\sigma)$, to a given optimal profile $\mathbf{m_*}$.  
We denote the family by $T(s) R(\sigma) \mathbf{m_*}$.  In polar coordinates,
$T(s) R(\sigma) \mathbf{m_*}$ is given by $\phi(x) = \sigma$ 
(the optimal profile lies in a fixed half-plane), and
$\theta(x) = \theta_*((x-s)/\Dwid)$, where $\Dwid = \sqrt{A/K_1}$ and
\begin{equation}
\label{eq: theta_*}
\theta_*(\xi) = 2 \tan^{-1}(e^{-\xi}).
\end{equation} 
It is clear that $\theta_*(\xi)$ satisfies
\begin{equation}
\label{eq: sin theta_* etc}
\theta_*' = -\sin\theta_*,
 \qquad   \sin \theta_*(\xi) = \sech \xi. 
 \end{equation}

 The dynamics of the magnetization in the presence of an applied
 magnetic field is described by the Landau-Lifschitz-Gilbert equation \cite{HubertSchaefer98},
 which for convenience we write in the equivalent Landau-Lifschitz
 (LL) form,
\begin{equation}\label{eq: LL}
\dot {\mathbf{m}} = \mathbf{m}\times (\mathbf{H } + \mathbf{H_a}) - \alpha \mathbf{m}\times \left(
\mathbf{m}\times (\mathbf{H } + \mathbf{H_a})
\right).
\end{equation}
Here $\alpha > 0$ is the  damping parameter, and  
we take the applied field  to lie along $\mathbf{\hat x}$,
\begin{equation}
\label{ eq: H_a}
\mathbf{H_a} = H_1(t) \mathbf{\hat x}.\end{equation}
In polar coordinates,
the LL equation is given by
\begin{align}
\label{eq: LL polar}
   \dot \theta &= \alpha e_1 - e_2 - \alpha H_1 \sin\theta,  \\
    \sin \theta \dot \phi &=   e_1 + \alpha e_2 - H_1\sin\theta.
\end{align}

The precessing solution is  a time-dependent translation and rotation of an optimal profile, which we write as $T(x_0(t)) R(\phi_0(t)) \mathbf{m_*}$. 
The centre $x_0(t)$ and orientation $\phi_0(t)$ of the domain wall for the precessing solution evolve according to
\begin{equation}
\label{eq: x_0(t) and phi_0(t)}
\dot x_0 = -\alpha\Dwid H_1, \quad \dot \phi_0 = -H_1.
\end{equation}
It  was shown~\cite{Sun10,our_PRL} that $T(x_0) R(\phi_0) \mathbf{m_*}$ satisfies the LL equation.
%

It is
important to note that the precessing solution
is fundamentally different
from the so-called Walker solution \cite{SchryerWalker74}. Indeed, the
latter is defined only for $K_2 > 0$ (the fully anisotropic case) and
time-independent $H_1$ less than the breakdown field $H_W = \alpha
K_2/2$. The Walker solution is given by $\m(x,t) = \big( \cos \theta_W
(x,t), \sin\theta_W (x,t) \cos\phi_W, \sin\theta_W (x,t) \sin\phi_W
\big)$ with
\begin{align}
  &\theta_W(x,t) = \theta_* \big( \gamma^{-1} (x - V_Wt) \big) \,, \label{eq:Wsolution1}\\
  &\sin 2\phi_W = H_1/H_W \,, \label{eq:Wsolution2}
\end{align}
and
\begin{align}
  & V_W = \gamma (\alpha+\alpha^{-1}) \Dwid 
  H_1 \,, \label{eq:Wsolution3}\\
  &\gamma = \left( \frac{K_1}{K_1 + K_2\cos^2\phi_W} \right)^{\frac{1}{2}}
  \,. \label{eq:Wsolution4}
\end{align}
Equations~(\ref{eq:Wsolution1})-(\ref{eq:Wsolution4}) describe a DW
traveling with a constant velocity $V_W$ whose magnitude cannot 
exceed $\gamma (\alpha+\alpha^{-1}) \Dwid H_W$; note that
$V_W$ does not depend linearly on the applied field $H_1$.  In contrast, the velocity $\dot x_0$ of the precessing solution is proportional to $H_1$, and can be arbitrarily large.  
Also, while for the Walker solution the plane of the DW remains
fixed, for the precessing solution it rotates about the nanowire at a
rate proportional to $H_1$. Finally, for the Walker solution, the DW
profile contracts ($\gamma < 1$) in response to the applied field,
whereas for the precessing solution the DW profile propagates without
distortion.  

In this paper we consider the stability of the precessing solution.  We establish linear stability  with
respect to perturbations of the initial optimal profile (Sec.~\ref{sec:profile}), small hard-axis anisotropy  (Sec.~\ref{sec:anisotropy}), and small transverse applied magnetic field (Sec.~\ref{sec:field}); specifically, we show, to leading order in the perturbation parameter,  that up to translation and rotation, the perturbed solution converges to the  precessing solution (in the case of perturbed initial conditions)  or   stays close to it for all times (for small  hard-axis anisotropy and small transverse magnetic field).  The argument is based on considerations of energy, and depends on the fact that for all $t$, the precessing solution belongs to the family of global minimizers.  The analytic argument
 establishes only linear stability.  Nonlinear stability is verified numerically for all three cases in Sec.~\ref{sec: num}.
 For convenience we choose units so that $A = K_1 = 1$.

\section{Perturbed initial profile}
\label{sec:profile}

Let ${\bf m_\epsilon}(x,t)$ denote the solution of the LL equation with  
initial condition $\mathbf{m_*} + \epsilon \boldsymbol {\mu}$,  a perturbation of an optimal profile. Let $T(x_\epsilon(t)) R(\phi_\epsilon(t)) \mathbf{m_*}$ denote the optimal profile which, at time $t$, is closest to ${\bf m_\epsilon}$; that is,  the quantity 
\begin{equation}
\label{ eq: norm difference}
|| \mathbf{m_\epsilon} - T(s) R(\sigma) \mathbf{m_*}||^2 =
\int  \big(\mathbf{m_\epsilon}(x,t) - R(\sigma)\mathbf{ m_*}(x-s)\big)^2\,dx
\end{equation} 
is minimized for $s =x_\epsilon(t)$ and $\sigma = \phi_\epsilon(t)$.  Then the following conditions must hold:
\begin{align}
\label{ eq: optimal xbar and phibar}
\int \mathbf{m_\epsilon}\cdot \left(T( x_\epsilon(t)) R(\phi_\epsilon(t))\frac{\partial \mathbf{m_*} }{\partial x} \right)\,dx &= 0,\nonumber\\
\int \mathbf{m_\epsilon} \cdot \left( \mathbf{\hat x} \times T( x_\epsilon(t)) R(\phi_\epsilon(t))\mathbf{ m_*} \right) \,dx &= 0.
\end{align}
It is clear that $ x_\epsilon(t) = x_0(t) + O(\epsilon)$ and $\phi_\epsilon(t) = \phi_0(t) + O(\epsilon)$, but we shall not explicitly calculate the $O(\epsilon)$ corrections produced by the perturbation.  Rather, our approach is to show that
to leading order $O(\epsilon^2)$,  
$||\mathbf{m_\epsilon} - T(x_\epsilon) R(\phi_\epsilon) \mathbf{m_*}||^2$ decays to zero with $t$.  This will imply that the precessing solution is linearly stable under perturbations of initial conditions up to translations and rotations.

Let $\theta_\epsilon(x,t)$ and $\phi_\epsilon(x,t)$ denote the
spherical coordinates of $\mathbf{m_\epsilon}(x,t)$.  We expand these
in an asymptotic series,
\begin{align}
\label{eq: perturbed polar solution}
    \theta_\epsilon(x,t) &= \theta_*(x - x_\epsilon(t)) &+ &\ \epsilon \theta_1( x-x_\epsilon(t),t) + \cdots, \nonumber \\
    \phi_\epsilon(x,t) &=  \phi_*(t)  &+  & \ \epsilon \phi_1( x-x_\epsilon(t),t) + \cdots
\end{align}
where the correction terms $\theta_1(\xi, t)$, $\phi_1(\xi,t)$, etc are expressed in  a reference frame  moving with the domain wall
Then  to leading order $O(\epsilon^2)$,
\begin{multline}
\label{ eq: norm diff in terms of polars}
|| \mathbf{m_\epsilon} - T(x_\epsilon) R(\phi_\epsilon) \mathbf{m_*}||^2 =
\epsilon^2 \int  ( \theta_1^2 + \sin^2\theta_* \phi_1^2) \, d\xi\\
= \epsilon^2\braket{\theta_1}{\theta_1} + \epsilon^2\braket{\sin\theta_*\phi_1}{\sin\theta_*\phi_1},
\end{multline}
where for later  convenience we have introduced  Dirac notation, expressing the integral in Eq.~\eqref{ eq: norm diff in terms of polars} in terms of inner products. 
It is straightforward to show that the conditions Eq.~\eqref{ eq: optimal xbar and phibar} imply (using $\theta_*' = -\sin\theta_*)$)  that
\begin{equation}
\label{eq: optimal x phi polar}
    \braket{\sin\theta_*}{\theta_1} =   \braket{\sin\theta_*}{\sin\theta_* \phi_1} =  0,
\end{equation}\\
which expresses the fact that the perturbations described by $\theta_1$ and $\phi_1$ are orthogonal to infinitesimal translations (described  by $\sin\theta_*$) along and rotations about $\mathbf{\hat x}$.


Since the difference between $\mathbf{m_\epsilon}$ and $T(x_\epsilon) R(\phi_\epsilon)
\mathbf{m_*}$ is  $O(\epsilon)$, the difference in their energies is  $O(\epsilon^2)$ (as $T(x_\epsilon) R(\phi_\epsilon)
\mathbf{m_*}$ satisfies the Euler-Lagrange equation Eq.~\eqref{eq: Euler-Lagrange}), and is given to leading order by the second variation of $E$ about $\mathbf{m_*}$,
\begin{gather}
\Delta E_\epsilon = E(\mathbf{m_\epsilon}) - E(T(x_\epsilon) R(\phi_\epsilon)
\mathbf{m_*}) =\nonumber \\E(\mathbf{m_\epsilon}) - E(
\mathbf{m_*}) = \frac{\epsilon^2}{2} \int f_0\, d\xi,\label{eq: energy difference} 
\\
\label{eq: f_0}
\text{where } f_0 = {\theta_1'}^2 + \cos 2 \theta_* \theta_1^2 + \sin^2 \theta_* {\phi_1'}^2.\nonumber
\end{gather}
Using the relations Eq.~\eqref{eq: sin theta_* etc} and performing some integrations by parts, we can write  
\begin{equation}
\label{eq: quadratic in terms of H}
\int f_0 \,d\xi = \me{\theta_1}{\mathcal{H}}{\theta_1} +   \me{\sin\theta_* \phi_1}{\mathcal{H}}{\sin\theta_*\phi_1},
\end{equation}
where $\mathcal{H}$ is the Schr\"odinger operator $-d^2/d\xi^2 + V(\xi)$ with potential given by
\begin{equation}
\label{ eq: Hcal}
V(\xi) = 1 - 2\sech^2\xi.
\end{equation}
$V(\xi)$ is a particular case of the P\"oschl-Teller potential, for which the spectrum of   $\mathcal{H}$ is known \cite{MorFesh}.  $\mathcal{H}$ has two eigenstates, namely  $\sin\theta_*(\xi)= \sech \xi$ with eigenvalue  $\lambda_0 = 0$, and $\cos\theta_*(\xi) = \tanh \xi$ with eigenvalue $\lambda_1 = 1$, and its continuous spectrum is bounded below by $\lambda = 1$.  This is consistent with the fact that the optimal profiles are global minimizers of  $E$  (subject to the boundary conditions Eq.~\eqref{ eq: bcs}), which implies that the second variation of $E$ about $\mathbf{m_*}$
is positive for variations transverse to   translations and rotations of $\mathbf{m_*}$.  It follows that, for any (smooth) square-integrable function $f(\xi)$ orthogonal to $\sin\theta_*$, 
we have  that
\begin{equation}
\label{ eq: general H inequality}
\me{f}{\mathcal{H}^{j+1}}{f} \ge \me{f}{\mathcal{H}^j}{f}
\end{equation}
for $j \ge 0$ (we will make use of this for $j=0$ and $j=1$).  In particular, 
since $\theta_1$ and $\sin\theta_* \phi_1$ are orthogonal to $\sin\theta_*$ (cf Eq.~\eqref{eq: optimal x phi polar}), it follows that
\begin{align}
\me{\theta_1}{\mathcal{H}}{\theta_1}
& 
\ge  \braket{\theta_1}{\theta_1},
\label{eq: spectral bound on Htheta_1} \\
 \me{\sin\theta_* \phi_1}{\mathcal{H}}{\sin\theta_*\phi_1}
 &
 \ge  \braket{\sin\theta_*\phi_1}{\sin\theta_* \phi_1}. \label{eq: spectral bound on H phi_1}
\end{align}
Therefore, from the preceding Eqs.~\eqref{eq: spectral bound on Htheta_1}--\eqref{eq: spectral bound on H phi_1}
and Eqs.~\eqref{ eq: norm diff in terms of polars}  and \eqref{eq: energy difference}--\eqref{eq: quadratic in terms of H}, we get, to leading order $O(\epsilon^2)$, that
\begin{equation}
\label{eq: m diff bounded by E diff second }
|| \mathbf{m_\epsilon} - T(x_\epsilon) R(\phi_\epsilon) \mathbf{m_*}||^2 \le
2 \Delta E_\epsilon .
\end{equation}


Below we show that, to leading order $O(\epsilon^2)$, for small enough $H_1$ (it turns out that $|H_1| < 1/2$ is sufficient), we have the inequality
\begin{equation}
\label{ eq: dE/dt inequality}
\frac{d}{dt}
\Delta E_\epsilon
\le -\gamma \Delta E_\epsilon
\end{equation}
for some $\gamma > 0$.
Taking Eq.~\eqref{ eq: dE/dt inequality} as given,
it follows from the Gronwall inequality that
\begin{equation}
\label{eq: gronwall 1 }
 \Delta E_\epsilon \le \frac{1}{2} C \epsilon^2 e^{-\gamma t}
\end{equation}
for some $C > 0$ (which depends only on the form of the initial perturbation).
From Eq.~\eqref{eq: m diff bounded by E diff second }, it follows that
\begin{equation}
\label{eq: m diff bounded by E diff third}
|| \mathbf{m_\epsilon} - T(x_\epsilon) R(\phi_\epsilon) \mathbf{m_*}||^2 \le
C \epsilon^2 e^{-\gamma t}.
\end{equation}

The result Eq.~\eqref{eq: m diff bounded by E diff third} shows that, to $O(\epsilon^2)$, $\mathbf{m_\epsilon} $
converges to an optimal profile with respect to the $L^2$-norm.  In fact, with a small extension of the argument,  we can also show that, to $O(\epsilon^2)$, $\mathbf{m_\epsilon} $ converges to an optimal profile uniformly (that is, with respect to the $L^\infty$-norm).   Indeed, making use of the preceding estimates, one can obtain a bound on $||\mathbf{m'_\epsilon} - T(x_\epsilon) R(\phi_\epsilon) \mathbf{m'_*}||$, the $L^2$-norm of  the difference in the spatial derivatives of the perturbed solution and the optimal profile.  To $O(\epsilon^2)$,
\begin{multline}
\label{eq: derivative bound }
||\mathbf{m'_\epsilon} - T(x_\epsilon) R(\phi_\epsilon) \mathbf{m'_*}||^2 \\= \epsilon^2\left(
\braket{\theta'_1}{\theta'_1} +  \braket{\sin\theta_*\phi'_1}{\sin\theta_*\phi'_1}  +
 \braket{\sin\theta_*\theta_1}{\sin\theta_*\theta_1}  
 \right)\\
 \le \epsilon^2 \left(3( \me{\theta_1}{\mathcal{H}}{\theta_1} +  \me{\sin\theta_* \phi_1}{\mathcal{H}}{\sin\theta_*\phi_1}\right)\\
 \le 6\epsilon^2 \Delta E_\epsilon.
\end{multline}
Arguing as in Eqs.~\eqref{eq: m diff bounded by E diff second }--\eqref{eq: m diff bounded by E diff third}, we may conclude that $||\mathbf{m'_\epsilon} - T(x_\epsilon) R(\phi_\epsilon) \mathbf{m'_*}||$ decays exponentially with $t$.
Thus, $\mathbf{m_\epsilon}$ converges to an optimal profile with respect to the Sobolev $H^1$-norm (where $||f||^2_{H^1} = ||f||^2 +  ||f'||^2$).   It is a standard result that this implies that the convergence is also uniform (again, to $O(\epsilon^2)$).

%
%
%
%
%

It remains to establish Eq.~\eqref{ eq: dE/dt inequality}. From Eq.~\eqref{eq: LL}, we have that for any solution $\mathbf{m}(x,t)$ of the LL equation,
\begin{multline}
\label{eq: energy rate 1 }
\frac{d}{dt}E(\mathbf{m})= 
-\int \mathbf{H}\cdot \mathbf{\dot m}\, dx \\
= \int\left( \mathbf{m}\times \mathbf{H}\right)\cdot\mathbf{H_a} \,dx -\\
- \alpha\int
\left(\mathbf{m}\times \mathbf{H})^2 + (\mathbf{m}\times \mathbf{H})\cdot (\mathbf{m}\times \mathbf{H_a}\right)\, dx\\
= -\alpha \int \left(e_1^2 + e_2^2 + H_1 \sin\theta e_1\right)\,dx,
\end{multline}
where $e_1$ and $e_2$ are given by Eq.~\eqref{eq: e_0 - e_2}, and we have used the fact that the term $(\mathbf{m}\times \mathbf{H})\cdot\mathbf{H_a}$  vanishes on integration.  Substituting the perturbed solution $\mathbf{m_\epsilon}$ into Eq.~\eqref{eq: energy rate 1 } and noting that the $E(T(x_\epsilon) R(\phi_\epsilon) \mathbf{m_*}) = E(\mathbf{m_*})$ does not vary in time, we obtain after some straightforward manipulation that
\begin{multline}
\label{ eq: energy rate m_*}
\frac{d}{dt} \Delta E_\epsilon
= \\ -\alpha \epsilon^2\big(
\me{\theta_1}{\mathcal{H}^2}{\theta_1} + 
\me{\sin\theta_* \phi_1}{\mathcal{H}^2}{\sin\theta_* \phi_1}
+  H_1 F\
\big)
\end{multline}
to leading $O(\epsilon^2)$,
where
\begin{equation}
\label{ eq: R}
F = \int \left(
\cos\theta_* f_0 + \cos\theta_*\sin^2\theta_* \theta_1^2
\right)\, d\xi.
\end{equation}
For the first two terms on the rhs of Eq.~\eqref{ eq: energy rate m_*}, we have, 
from Eq.~\eqref{ eq: general H inequality} and Eqs.~\eqref{eq: energy difference}--\eqref{eq: quadratic in terms of H}, that
\begin{multline}
\label{ eq: H^2 bound}
\me{\theta_1}{\mathcal{H}^2}{\theta_1} + 
\me{\sin\theta_* \phi_1}{\mathcal{H}^2}{\sin\theta_* \phi_1}\\  \ge
\me{\theta_1}{\mathcal{H}}{\theta_1} + 
\me{\sin\theta_* \phi_1}{\mathcal{H}}{\sin\theta_* \phi_1}\\
=\frac{2}{\epsilon^2} \Delta E_\epsilon.
\end{multline}
The term  $H_1 F$ in Eq.~\eqref{ eq: energy rate m_*} is not necessarily positive, as $H_1$ can have arbitrary sign. But for sufficiently small $|H_1|$, it is smaller in magnitude than the preceding two terms.  Indeed, we have, again using Eq.~\eqref{ eq: general H inequality}  and Eqs.~\eqref{eq: energy difference}--\eqref{eq: quadratic in terms of H}, that
\begin{multline}
\label{ eq: R estimate}
|F| \le \int \left(|f_0| + {\theta_1}^2\right)\,d\xi \le 
\frac{2}{\epsilon^2} \Delta E_\epsilon + \braket{\theta_1}{\theta_1} \\
\le \frac{2}{\epsilon^2} \Delta E_\epsilon + \me{\theta_1}{\mathcal{H}}{\theta_1} 
\le\frac{4}{\epsilon^2} \Delta E_\epsilon.
\end{multline}
Substituting 
Eqs.~\eqref{ eq: H^2 bound} and \eqref{ eq: R estimate} into Eq.~\eqref{ eq: energy rate m_*}, we get that
\begin{equation}
\label{ eq: rate inequality final}
\frac{d}{dt} \Delta E_\epsilon
\le -2\alpha (1 - 2|H_1|) \Delta E_\epsilon,
\end{equation}
from which the required estimate \eqref{ eq: dE/dt inequality}  follows for $|H_1| < 1/2$.

It is to be expected that the stability of the precessing solution depends on the applied field not being   too large.  Indeed, it is easily shown that, for $H_1 > 1$ (resp.~$H_1 < -1$), the static, uniform solution $\mathbf{m}
 = -\mathbf{\hat x}$ (resp.~ $\mathbf{m}
 = +\mathbf{\hat x}$) becomes linearly unstable.  As the precessing solution is nearly uniform away from the domain wall, one would expect it to be similarly unstable for $|H_1| > 1$. The numerical results of Sec.~\ref{sec:num:profile} bear this out.  Finally, we remark that the stability criterion obtained here, namely $|H_1| < 1/2$, is certainly not optimal.

\section{Small hard-axis anisotropy}
\label{sec:anisotropy}
Next we suppose the hard-axis anisotropy is small but nonvanishing,
taking $K_2 = \epsilon  > 0$.  Let $\mathbf{m_\epsilon}(x,t)$ denote the solution of the LL equation with initial condition $\mathbf{m_\epsilon}(x,0) = \mathbf{m_*}(x)$.  As above, let $T(x_\epsilon(t)) R(\phi_\epsilon(t)) \mathbf{m_*}$ denote the translated and rotated optimal profile closest to $\mathbf{m_\epsilon}$ at time $t$.   Adapting the argument of the preceding section, we show below that, to leading order $O(\epsilon^2)$,
\begin{equation}
\label{eq: m diff bounded by E diff K_2 not zero}
|| \mathbf{m_\epsilon} - T(x_\epsilon) R(\phi_\epsilon) \mathbf{m_*}||^2 \le
C_2 \epsilon^2  \text{ for all $t > 0$}
\end{equation}
for some constant $C_2 > 0$.  In contrast to the preceding result Eq.~\eqref{eq: m diff bounded by E diff third} for perturbed initial conditions, here we do not expect $\mathbf{m_\epsilon}$ to converge to $T(x_\epsilon) R(\phi_\epsilon) \mathbf{m_*}$.  Indeed, while an explicit analytic solution of the LL equation is not available for small $K_2 $ (the Walker solution is valid only for $K_2 > 2 |H_1|/\alpha$), it is easily verified that there are no exact solutions of the form  $T(x_\epsilon(t)) R(\phi_\epsilon(t)) \mathbf{m_*}$.  The result Eq.~\eqref{eq: m diff bounded by E diff K_2 not zero} demonstrates that, through linear order in $\epsilon$, the solution for $K_2 = \epsilon $ remains close to the precessing solution, up to translation and rotation.

To proceed, let $\Delta E_\epsilon$ denote, as above, the difference
in the {\it uniaxial} micromagnetic energy, i.e.~ the energy given by
Eq.~\eqref{ eq: E} with $K_2 = 0$, between $\mathbf{m_\epsilon}$ and
$T(x_\epsilon) R(\phi_\epsilon) \mathbf{m_*}$.  Then, as in
Eq.~\eqref{eq: m diff bounded by E diff second }, we have that
\begin{equation}
\label{eq: m diff bounded by E diff second K_2 > 0 }
|| \mathbf{m_\epsilon} - T(x_\epsilon) R(\phi_\epsilon) \mathbf{m_*}||^2 \le
2\Delta E_\epsilon.
\end{equation}
As $E(T(x_\epsilon) R(\phi_\epsilon) \mathbf{m_*}) = E(\mathbf{m_*})$ is constant in time, we have that
\begin{equation}
\label{eq:  d Delta E/dt 1 K_2 > 0}
\frac{d}{dt} \Delta E_\epsilon 
=\frac{d}{dt} E(\mathbf{m_\epsilon}) .
\end{equation}

The hard-axis anisotropy affects  the rate of change of the uniaxial energy through additional terms in $\mathbf{\dot m}$.
Indeed, for any solution $\mathbf{m}(x,t)$ of the LL equation, we have that
\begin{equation}
\label{eq: energy rate 1 K_2 > 0 }
\frac{d}{dt} E(\mathbf{m})  =
\left. \frac{d}{dt}  \right|_{K_2 = 0}  E(\mathbf{m}) + G(\mathbf{m}),
\end{equation}
where $d/dt|_{K_2 = 0} E(\mathbf{m})$ denotes the rate of change when $K_2 = 0$, as given by Eq.~\eqref{eq: energy rate 1 },  and
\begin{multline}\label{eq: G} G(\mathbf{m}) =
-\e \int_\R (\m \cdot \yhat) (\m \times \field (\m))\cdot \yhat  \, \ud x  \\ +
\e \alpha \int (\m \times \field (\m)) \cdot (\m \times \yhat) (\m \cdot \yhat)  \, \ud x.
\end{multline}
Taking $\mathbf{m} = \mathbf{m_\epsilon}$, we recall from the preceding section (c.f.~Eq.~\eqref{ eq: dE/dt inequality}) that, for $|H_1| < 1/2$, 
\begin{equation}
\label{ eq: dE/dt inequality K_2 > 0}
\left. \frac{d}{dt}  \right|_{K_2 = 0}
E(\mathbf{m_\epsilon}) 
\le -\gamma \Delta E_\epsilon
\end{equation}
for some $\gamma > 0$.
Below we show that there exists constants $C_1, \gamma_1$ with $\gamma_1 < \gamma$ such that
\begin{equation}
\label{ eq: G inequality}
|G(\mathbf{m_\epsilon})| \le \gamma_1 \Delta E_\epsilon +  C_1 \epsilon^2  .
\end{equation} 
Taking Eq.~\eqref{ eq: G inequality} as given and substituting it along with Eq.~\eqref{ eq: dE/dt inequality K_2 > 0} into  Eqs.~\eqref{eq:  d Delta E/dt 1 K_2 > 0}--\eqref{eq: energy rate 1 K_2 > 0 }, we get that
\begin{equation}
\label{eq: dE/dt final inequality K_2 > 0 }
\frac{d}{dt} \Delta E_\epsilon \le -(\gamma - \gamma_1) \Delta E_\epsilon + C_1\epsilon^2 .
\end{equation}
From Gronwall's equality it follows that
\begin{equation}
\label{eq: Gronwall }
\Delta E_\epsilon \le \frac{C_1}{\gamma-\gamma_1} \epsilon^2 ,
\end{equation}
which together with Eq.~\eqref{eq: m diff bounded by E diff second K_2 > 0 } yields  the required result Eq.~\eqref{eq: m diff bounded by E diff K_2 not zero}.

It remains to show Eq.~\eqref{ eq: G inequality}.
Substituting the asymptotic expansion Eq.~\eqref{eq: perturbed polar solution}, we obtain after  straightforward calculations that,  to leading order $O(\epsilon^2)$,
\begin{multline} \label{eq:G(m_eps)}
G(\mathbf{m_\epsilon}) =
- \epsilon^2 \cos^2 \phi_* (t)\\ \times \int \left(\sin^4 \theta_* \phi_1'   + 4/3 \alpha \sin^3 \theta_*\theta_1'\right)\, d\xi.
\end{multline}
This can be estimated using the elementary inequality
\beq \label{eq:ineq}
2 |ab| \leq \beta a^2  + \frac{ b^2}{\beta},
\eeq
which holds for any $\beta > 0$.  Indeed, recalling Eqs.~\eqref{eq: sin theta_* etc}, \eqref{eq:  energy difference}, \eqref{eq: spectral bound on Htheta_1}, and using integration by parts where necessary, we have that
\begin{align}
\label{eq: inequality examples}
\left|\int  \sin^4 \theta_* \phi_1' \,d\xi\right|
&\le \frac{\beta}{2}\int \sin^2\theta_* {\phi_1'}^2\, d\xi + \frac{1}{2\beta} \int \sin^6\theta_* \, d\xi\nonumber\\
&\le \frac{\beta}{ \epsilon^2} \Delta E_\epsilon + \frac{8}{15\beta},\nonumber\\
\left| \int  \sin^3 \theta_*\theta_1' \,d\xi \right| &\le \frac{\beta}{2}\int {\theta'_1}^2\, d\xi + \frac{1}{2\beta} \int \sin^6\theta_* \, d\xi\nonumber\\
&\le \frac{\beta}{ \epsilon^2} \Delta E_\epsilon + \frac{8}{15\beta}.
\end{align}
From Eqs.~\eqref{eq:G(m_eps)}--\eqref{eq: inequality examples}, it is clear that $\beta$, $\gamma_1$ and $C_1$ can be chosen so that Eq.~\eqref{ eq: G inequality} is satisfied.

\section{Small transverse applied field}
\label{sec:field}
Suppose the applied magnetic field  has a small transverse component, so that
$\mathbf{H_a} = H_1 \mathbf{\hat x} +  H_2 \mathbf{\hat y}$,
where
\begin{equation}
\label{ eq: H_2}
H_2 = \epsilon h_2(x)
\end{equation}
($h_2$ depends on $x$ but not $t$).  For simplicity, let $K_2 = 0$.  Let $\mathbf{m_\epsilon}(x,t)$ denote the solution of the LL equation with initial condition $\mathbf{m_\epsilon}(x,0) = \mathbf{m_*}(x)$.  As above, let $T(x_\epsilon(t)) R(\phi_\epsilon(t)) \mathbf{m_*}$ denote the translated and rotated optimal profile closest to $\mathbf{m_\epsilon}$ at time $t$.
  
We first note that, unless $h_2$ vanishes as $x \rightarrow \pm \infty$, $\mathbf{m_\epsilon}$ will {\it not} remain close to $T(x_\epsilon(t)) R(\phi_\epsilon(t)) \mathbf{m_*}$.  For example, if $h_2$ is constant, then away from the domain wall, $\mathbf{m_\epsilon}$ will relax to one of the local minimizers of the homogeneous energy
$K_1 (1 - m_1^2) - \mathbf{H_a}\cdot\mathbf{m}$, and these do not lie  along $\pm \mathbf{\xhat}$ for $H_2 \neq 0$.  It follows that $||\mathbf{m_\epsilon} - T(x_\epsilon(t)) R(\phi_\epsilon(t)) \mathbf{m_*}||$ will diverge with time.  

Physically, this divergence is spurious.  It stems from the fact that we are taking the wire to be  of infinite extent.  One way to resolve the issue, of course, would be to take the wire to be of finite length.  However,   one would then no longer have an explicit analytic solution of the LL equation.

Here we shall take a simpler approach, and assume that the transverse field $h_2(x)$ approaches zero as $x$ approaches $\pm \infty$.  In fact, for technical reasons, it will be convenient to assume that 
the integral of $h_2^2 + {h_2'}^2$, i.e.~the squared Sobolev norm $||h_2||_{H^1}$, is finite.  Then without loss of generality, we may assume
\begin{equation}
\label{ eq: finite Sobolev norm of h_2}
||h_2||_{H^1}^2 = \int (h_2^2 + {h_2'}^2)\, d\xi = 1.
\end{equation}
Under this assumption, the main result of this section is that $\mathbf{m_\epsilon}$ stays close
to an optimal profile up to translation and rotation.  That is, for some $C_1 > 0$,
\begin{equation}
\label{eq: main result H_2 not zero}
|| \mathbf{m_\epsilon} - T(x_\epsilon) R(\phi_\epsilon) \mathbf{m_*}||^2 \le
C_1 \epsilon^2.
\end{equation}

The demonstration proceeds as in the preceding section, so we will discuss only the points at which  the present case is different.  The main difference is that, in place of Eq.~\eqref{eq:G(m_eps)}, we get (by considering the LL equation with $H_2 \neq 0$ rather than $K_2 \neq 0)$ the following expression for $G(\mathbf{m_\epsilon})$ to leading order $O(\epsilon^2)$:
\begin{multline} \label{eq:energy-time2}
G(\mathbf{m_\epsilon}) =\epsilon^2\Big( \alpha \cos \phi_*(t) \int \cos\theta_* \left( \theta_1'' -\cos 2\theta_* \theta_1 \right) h_2 \, d\xi \\
-\alpha \sin \phi_* (t) \int \sin\theta_*\left( \phi_1''  - 2 \cos \theta_* \phi_1'  \right) h_2 \, d\xi\\
-  \sin \phi_*(t) \int \left( \theta_1'' -\cos 2\theta_* \theta_1 \right)  h_2 \, d\xi\\
-  \cos \phi_* (t) \int
\sin\theta_*\cos\theta_* \left( \phi_1''  - 2 \cos \theta_* \phi_1'  \right) h_2 \, d\xi\Big).
\end{multline}
After some straightforward manipulations including integration by parts, and making use of  the inequality Eq.~\eqref{eq:ineq}, one can show that
\begin{align} \label{eq: estimates for G}
&\left| \int \cos\theta_* \left( \theta_1'' -\cos 2\theta_* \theta_1 \right) h_2 \, d\xi \right|
\leq {\beta \over 2} \| \theta_1 \|_{H^1}^2 + {1 \over 2 \beta}, \nonumber  \\
& \left| \int \sin\theta_*\left( \phi_1''  - 2 \cos \theta_* \phi_1'  \right) h_2 \, d\xi \right| \leq  {\beta \over 2} 
||\sin \theta_* \phi_1'||^2  + {1 \over 2 \beta} ,  \nonumber\\
&\left| \int  \left( \theta_1'' -\cos 2\theta_* \theta_1 \right)  h_2 \, d\xi \right| \leq {\beta \over 2} \| \theta_1 \|_{H^1}^2 + {1 \over 2 \beta} , \nonumber\\
&\left|  \int 
\sin\theta_*\cos\theta_* \left( \phi_1''  - 2 \cos \theta_* \phi_1'  \right) h_2  \, d\xi \right| \nonumber \\
&\qquad \qquad \qquad \qquad \qquad \leq {\beta \over 2}||\sin \theta_* \phi_1'||^2 + {1 \over 2 \beta}.
\end{align}
From Eqs.~\eqref{eq: energy difference}, \eqref{eq: quadratic in terms of H} and \eqref{eq: spectral bound on Htheta_1} it follows that
\begin{equation}
\label{eq: estimates for ||theta'_1||, || sin theta phi'||}
   \int \left({\theta_1'}^2 + \sin^2\theta_* {\phi_1'}^2\right)\,d\xi  \le \frac{4}{\epsilon^2} \Delta E_{\epsilon},
\end{equation}
and 
\begin{equation} \label{eq: estimate for ||theta||}
      \int \theta_1^2 \,d\xi  \le \frac{2}{\epsilon^2} \Delta E_{\epsilon}.\end{equation}
Substituting Eqs.~\eqref{eq: estimates for G}--\eqref{eq: estimate for ||theta||} into Eq.~\eqref{eq:energy-time2}, we get that
\begin{equation}
\label{eq: G ineq H transverse }
|G(\mathbf{m_\epsilon})| \le(1+\alpha) \left(3\beta \Delta E_\epsilon + \frac{1}{\beta}\epsilon^2 \right).
\end{equation}
This estimate is of the same form as \eqref{ eq: G inequality}, and the argument given there,  with $\beta$ chosen appropriately, establishes Eq.~\eqref{eq: main result H_2 not zero}.


\section{Numerical studies}\label{sec: num}

In the preceding Sections~\ref{sec:profile}--\ref{sec:field} we have
shown that the precessing solution is linearly stable; to leading
order $O(\epsilon)$, a perturbed solution either approaches or stays
close to the precessing solution up to a translation and rotation,
according to whether the perturbation is to the initial conditions or
to the anistropy and transverse applied magnetic field in the LL
equation.  Here we present numerical results which verify nonlinear
stability for the precessing solution under small perturbations.  To
this end, we investigate the energy, $\Delta E_\epsilon = E(
\mathbf{m_\epsilon}) - E(\mathbf{m_*})$, of the numerically computed
perturbed DW, $\mathbf{m_\epsilon}(x,t)$, relative to the minimum
energy $E(\mathbf{m_*})$ of an optimal profile, as a function of time
$t$.  Throughout, $E$ is taken to be the {\it uniaxial} micromagnetic
energy given by Eq.~\eqref{ eq: E} with  $K_2 = 0$.  As in the
preceding sections, we choose units so that $A = K_1 = 1$.  In these
units, $E(\mathbf{m_*}) = 2$. 
In typical ferromagnetic
microstructures, the value of the Gilbert damping parameter $\alpha$
is known to lie between 0.04 and 0.22 (see~e.g.~Ref.~\cite{Tserkovnyak} and references within), so we take $\alpha =
0.1$ throughout our numerical study.

\subsection{Perturbed initial profile}
\label{sec:num:profile}

We first investigate the evolution of a DW,
$\mathbf{m_\epsilon}(x,t)$, from an initial perturbation of an optimal
profile.  We take the initial condition in polar coordinates to be
given by
\begin{equation}
  \theta_{\epsilon} (x,0) = \theta_* \left(\frac{x}{1 + \epsilon_1}\right) \,, 
  \quad \phi_{\epsilon} (x) = \phi_0 + \epsilon_2 x \,,
\label{eq:num-02}
\end{equation}
which corresponds to stretching the unperturbed profile along and
twisting it around the axis of the nanowire. The applied field is
directed along the nanowire, $\field_a = H_1 \xhat$, and we take $K_2
= 0$.

\begin{figure}[ht]
\centerline{\epsfig{figure=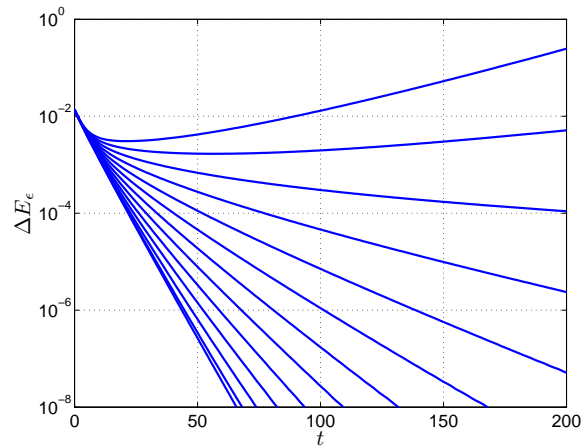,width=3in}}
\caption{(Color online) Relative energy, $\Delta E_\epsilon(t)$, of the
  perturbed DW for 13 different values of the applied field $H_1$. See
  text for discussion.}
\label{fig1}
\end{figure}

Figure~\ref{fig1} shows the dependence of the relative energy $\Delta
E_\epsilon$ on time $t$ for different values of the applied field
$H_1$. The figure presents 13 curves corresponding, from top to
bottom, to $H_1$ varying from $-1.2$ to $0$ at the increment of
$0.1$. In the initial condition given by Eq.~(\ref{eq:num-02}), we
take $\epsilon_1 = 0.1$ and $\epsilon_2 = \pi/50$.

Figure~\ref{fig1} clearly indicates that $\Delta E_\epsilon(t)$ decays
exponentially for weak applied fields, $|H_1| \leq 1/2$, in accord
with the analytic result Eq.~\eqref{eq: gronwall 1 }. However, for
$|H_1| \sim 1$, deviations from exponential decay are evident, and the
precessing solution appears to become unstable for $|H_1| \gtrsim 1$.

\subsection{Small hard-axis anisotropy}
\label{sec:num:K2}

We consider next the evolution of a DW from an optimal profile at $t =
0$ when the hard-axis anisotropy $K_2$ is nonvanishing. We fix $H_1 =
-0.5$.

\begin{figure}[ht]
\centerline{\epsfig{figure=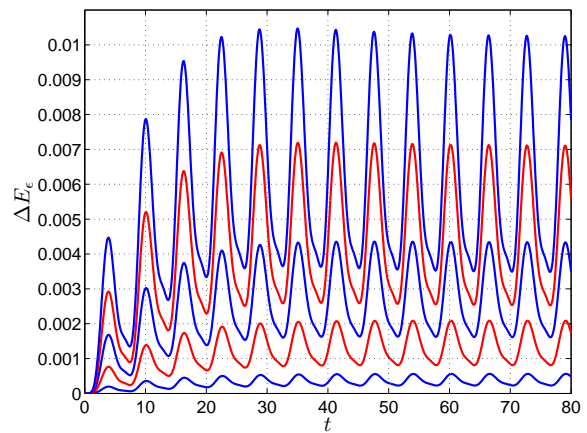,width=3in}}
\caption{(Color online) Relative energy, $\Delta E_\epsilon(t)$, of the
  perturbed DW for 5 different values of the hard-axis anisotropy
  constant $K_2$. See text for discussion.}
\label{fig2}
\end{figure}

Figure~\ref{fig2} shows the dependence of the relative energy $\Delta
E_\epsilon$ on time $t$ for different values of $K_2$. The figure
presents 5 curves corresponding, from top to bottom, to $K_2$ varying
from $0.1$ to $0.02$ at the decrement of $0.02$. (The blue and red
colorings alternate to make adjacent curves more easily
distinguishable.) It is evident that the relative energy remains
small, verifying the linear analysis of Sec.~\ref{sec:anisotropy}.

\begin{figure}[ht]
\centerline{\epsfig{figure=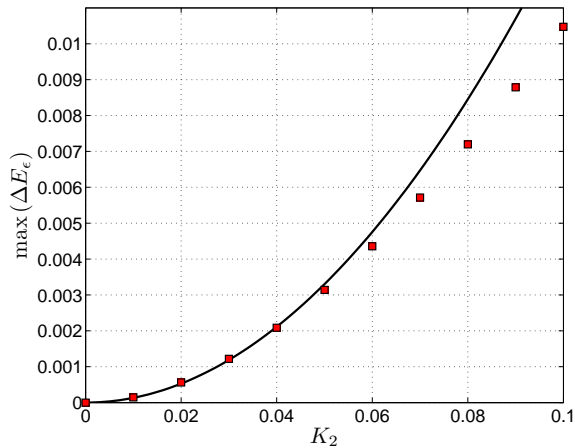,width=3in}}
\caption{(Color online) Maximum value of the relative energy $\Delta
  E_\epsilon$ of the perturbed DW as a function of the hard-axis
  anisotropy $K_2$. Numerically computed values are represented by
  (red) squares. The (black) solid curve is a parabola, $\max(\Delta
  E_\epsilon) = C_K K_2^2$ with $C_K = 1.3207$, fitted by the method
  of least squares through the data points with $K_2 \leq 0.04$.}
\label{fig3}
\end{figure}

Figure~\ref{fig3} shows the maximum value of the relative energy
$\Delta E_\epsilon$ (over the interval $0 \le t \le 80$) as a function
of $K_2$. Red squares represent numerically computed values. The black
solid curve is the parabola $ C_K K_2^2$, with $C_K = 1.3207$ fitted
by the method of least squares through the data points with $K_2 \leq
0.04$. We obtain convincing confirmation of the leading-order
analytical result Eq.~\eqref{eq: Gronwall }.  For larger values of
$K_2$, we see departures from quadratic dependence; for sufficiently
large values of $K_2$ (not shown), the Walker solution was recovered.

\subsection{Small transverse applied field}
\label{sec:num:H2}

Finally, we address the stability of the precessing solution under an
applied magnetic field, $\field_a = H_1 \xhat + H_2 \yhat$, with a
small transverse component, $H_2(x)$.  As discussed in
Sec.~\ref{sec:field}, we want $H_2(x)$ to vanish as $x\rightarrow
\pm\infty$. Here we take
\begin{equation}
\label{ eq: H_2 window}
H_2(x) = {\bar H}_{2} w(x),
\end{equation}
where $w(x)$ is equal to one inside the window $0 \le x \le 20$ and
vanishes outside (the argument of Section~\ref{sec:field} is easily
modified to establish the linear stability result Eq.~\eqref{eq: Gronwall } in this case). We
consider the evolution of a DW given at $t = 0$ by the optimal profile
$\mathbf{m_*}$ centred at $x=0$.  We take $H_1 = -0.5$, so that in the
absence of the transverse field, the DW velocity is positive
(cf.~Eq.~\eqref{eq: x_0(t) and phi_0(t)}) and the DW crosses the
window. We take $K_2 = 0$.

\begin{figure}[ht]
\centerline{\epsfig{figure=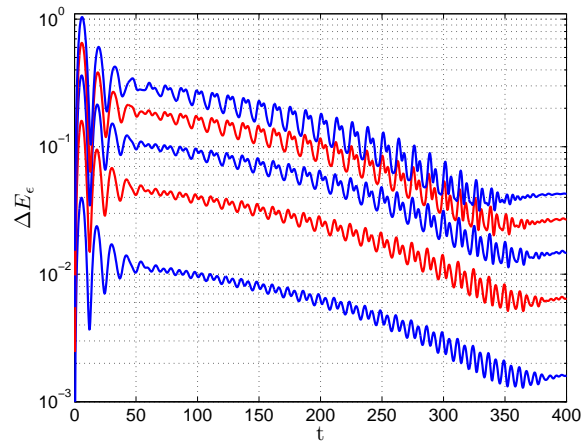,width=3in}}
\caption{(Color online) Relative energy, $\Delta E_\epsilon(t)$, of
  the perturbed DW for 5 different values of the transverse field
  amplitude $\bar H_2$. See text for discussion.}
\label{fig4}
\end{figure}

Figure~\ref{fig4} shows the dependence of the relative energy $\Delta
E_\epsilon$ on time $t$ for different values of the transverse field
amplitude $\bar H_2$. The figure presents 5 curves corresponding, from
top to bottom, to $\bar H_2$ varying from $0.1$ to $0.02$ at the
decrement of $0.02$. (The blue and red colorings alternate to make
adjacent curves more easily distinguishable.) The relative energy
$\Delta E_\epsilon(t)$ is presented over the time interval $0 \leq t
\leq 400$, which, for small values of $\bar H_2$, is sufficient for
the DW to traverse the spatial window $0 \le x \le 20$
(cf.~Eq.~(\ref{eq: x_0(t) and phi_0(t)})). The results confirm that
the relative energy of the perturbed magnetization profile remains
small for small values of $\bar H_2$, in accord with the leading-order
results of Section~\ref{sec:field}.

\begin{figure}[ht]
\centerline{\epsfig{figure=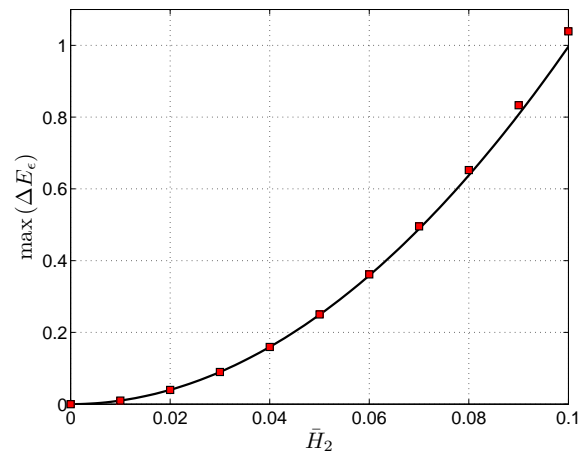,width=3in}}
\caption{(Color online) Maximum value of the relative energy $\Delta
  E_\epsilon$ of the perturbed DW as a function of the amplitude of
  the transverse applied field, $\bar H_2$. Numerically computed
  values are represented by (red) squares. The (black) solid curve is
  a parabola, $\max(\Delta E_\epsilon) = C_H \bar H_2^2$ with $C_H =
  99.6586$, fitted by the method of least squares through the data
  points with $\bar H_2 \leq 0.04$.}
\label{fig5}
\end{figure}

Figure~\ref{fig5} shows the maximum value of the relative energy
$\Delta E_\epsilon$ (over the interval $0 \le t \le 400$) as a
function of $\bar H_2$. Red squares represent numerically computed
values. The black solid curve corresponds to the parabola $C_H \bar
H_2^2$ with $C_H = 99.6586$ fitted by the method of least squares
through the data points with $\bar H_2 \leq 0.04$. The figure provides
a confirmation of the leading-order analytical result of
Sec.~\ref{sec:field} that the maximum relative energy depends
quadratically on $\bar H_2$ for small $\bar H_2$. Deviations from the
parabolic dependence can be seen for $\bar H_2 \gtrsim 0.08$.


\section{Conclusions}

The precessing solution is a new, recently reported exact solution of the Landau-Lifschitz-Gilbert equation.  It describes the evolution of a magnetic domain wall in a one-dimensional wire with uniaxial anisotropy subject to a spatially uniform but time-varying applied magnetic field along the wire.  We have analysed the stability of the precessing solution.  We have proved linear stability with respect to small perturbations of the initial conditions as well as to small hard-axis anisotropy and  small transverse applied fields, provided the applied magnetic field along the wire is not too large. We have also carried out numerical calculations that confirm full nonlinear stability under these perturbations.

Numerical calculations suggest that, for sufficiently large perturbations and applied longitudinal fields, the precessing solution becomes unstable, and new stable solutions appear.  It would be interesting to analyse these bifurcations and study these new regimes for DW motion.


\end{document}